\documentclass[a4paper,11pt]{article}
\usepackage{pos}
\usepackage{float}
\usepackage{tikz}
\usepackage{gensymb}
\usepackage{booktabs}
\usepackage[percent]{overpic}
\usetikzlibrary{shapes.geometric, arrows.meta, positioning}
\tikzstyle{layer} = [
    rectangle, draw=black, rounded corners,
    minimum width=1.5cm, minimum height=1.2cm,
    text centered, align=center
]
\tikzstyle{arrow} = [->, thick]

\title{Machine learning pipeline for identifying tracks of muons and hadrons at GRAPES-3 muon telescope}
 \ShortTitle{ML pipeline for identifying tracks of muons and hadrons at GRAPES-3 muon telescope}

\author*[a]{Mansi Talwar}
\author[a]{Sambit Sarkar}
\author[a]{Pravata K. Mohanty}
\onbehalf{on behalf of the GRAPES-3 collaboration}

\affiliation[a]{Tata Institute of Fundamental Research,\\
  Homi Bhabha Road, Mumbai 400005, India}

\emailAdd{mansi.talwar@tifr.res.in}
\emailAdd{pkm@tifr.res.in}

\abstract{The GRAPES-3 experiment is a ground-based extensive air shower array which consists of approximately $400$ closely packed plastic scintillator detectors and a large area muon telescope. Estimating the number of associated muons created in an air shower is crucial to understand the properties of primary cosmic rays. The GRAPES-3 muon telescope (G3MT) records these secondary muons, however, the punch-through hadrons can introduce background noise. This study aims to develop a machine learning pipeline to distinguish the tracks of secondary muons and hadrons at G3MT. We have used CORSIKA-simulated proton showers having energy in the range 100–158 TeV as an input for a Geant4-based detector simulation to analyze the signatures of both type of particles. Initially, single-particle classification was performed using decision trees, random forests, neural networks, and XGBoost, with XGBoost achieving the highest accuracy of 88.7\%. For multiparticle classification, we modelled Graph Neural Networks (GNNs) where each event was represented as a graph with detector hits as nodes. A GNN with edge convolution layers was developed to classify each node as a muon or hadron hit. Following this, a deep learning regression model using Dynamic Reduction Network was developed to estimate the number of particles and muons striking G3MT simultaneously. Details of the analysis and results of the multiparticle classification task will be presented.}

\ConferenceLogo{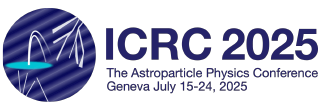}

\FullConference{39th International Cosmic Ray Conference (ICRC2025)\\
 15–24 July 2025\\
Geneva, Switzerland\\}

\begin{document}
\maketitle

\section{GRAPES-3 Experiment}
\begin{figure}[H]
    \centering
    \begin{minipage}{0.40\textwidth}  
        \centering
        \includegraphics[width=\textwidth]{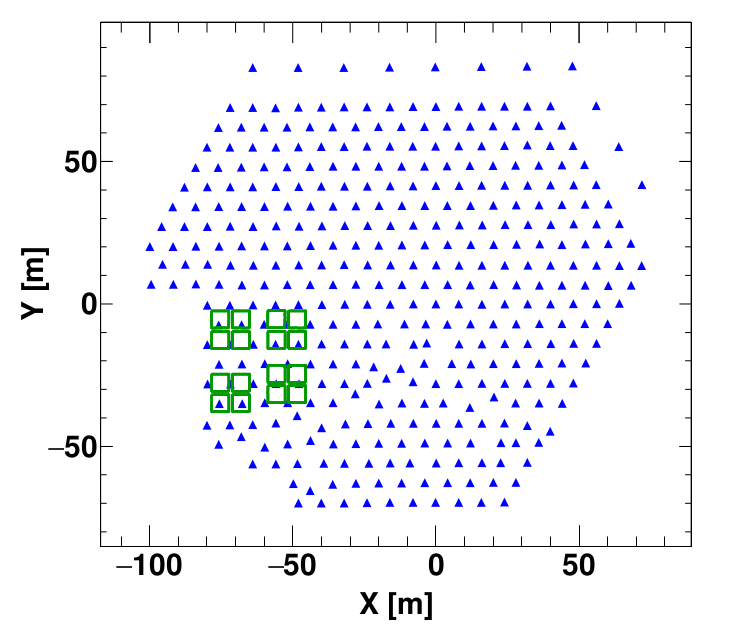} 
    \end{minipage} \hspace{0.05\textwidth}  
    \begin{minipage}{0.5\textwidth}  
        \centering
        \includegraphics[width=\textwidth]{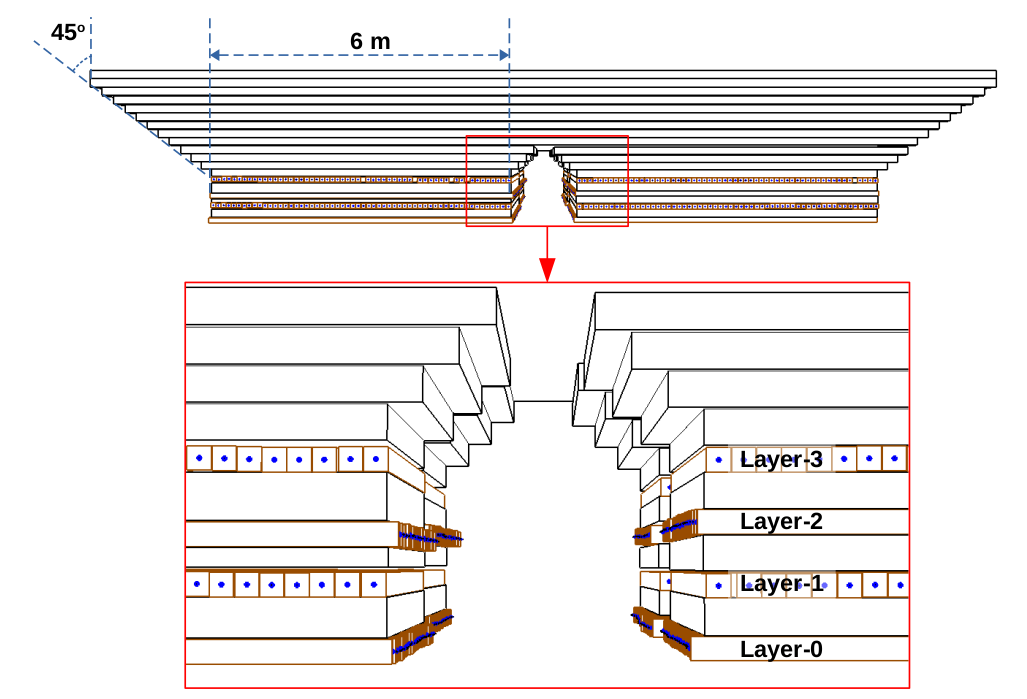}  %
    \end{minipage}
    \caption{Left:A schematic of GRAPES-3 experiment site showing hexagonal arrays of Scintillator detectors(shown in blue triangles) and Muon telescope(shown by green squares). Right: Zoomed in view of two adjacent muon modules having 4 layers of proportional counters in orthogonal alignment, each separated by concrete blocks\cite{varsi2023}.}
    \label{grapes-schematic}
\end{figure}

The GRAPES-3 (Gamma Ray Astronomy at PeV EnergieS – phase 3) is a ground-based extensive air shower (EAS) observatory located at Ooty(2200 m altitude, $11.4\degree$ N latitude and $76.7\degree$ E longitude) India to study the origin, acceleration and propagation of cosmic rays in Tera to Peta electronVolt energies. The experimental setup consists of a dense hexagonal array of 400 plastic scintillator detectors, each covering an area of $1\,\mathrm{m}^2$ and are spaced $8\,\mathrm{m}$ apart. These scintillators measure the radial density of electron component of shower and gives an estimate of total number of electrons which helps in further estimating the energy of the primary particle. The electron component is also important because it generates trigger information about arrival and helps to determine the direction of arrival of the shower \cite{gupta2005}. The figure \ref{grapes-schematic} in the left illustrates a schematic layout of the GRAPES-3 EAS array. The 16 independent muon modules are represented by open green squares, the four adjacent modules are built under a single roof referred to as a super module, and four such super modules make up the muon telescope.

Furthermore, the experiment features a $560\,\mathrm{m}^2$ muon telescope(G3MT) consisting of 16 muon modules represented by green squares in left figure \ref{grapes-schematic} equipped with 3712 proportional counters (PRCs) to detect the muon component of EAS.
 Each module consists of four layers of Proportional counters(PRCs) separated by concrete blocks, having an effective detection area of $35\,\mathrm{m}^2$. The bottom-most layer is designated as layer-0, while the top-most layer is layer-3 as shown in the figure on the right\ref{grapes-schematic}. When a particle strikes the muon module, it leaves a number of hits in the stacked layers of proportional counters, through which we can determine the direction of incoming particle. The proportional counters in G3MT consist of a galvanized mild-steel tube that is $600\,\mathrm{cm}$ long, with a wall thickness of $0.23\,\mathrm{cm}$ and a square cross-section of $10\,\mathrm{cm} \times 10\,\mathrm{cm}$. Both ends of the PRC tube are sealed with flat mild steel plates of $0.6\,\mathrm{cm}$ thickness. A tungsten wire with a $100\,\mu\mathrm{m}$ diameter is mounted along the axis of the PRC, acting as the anode, while the steel body of the PRC serves as the cathode. The PRCs are filled with P-10 gas, which is a mixture of 90\% argon and 10\% methane, maintained at a pressure slightly higher than the local atmospheric pressure in Ooty. Concrete blocks, each measuring $60\,\mathrm{cm} \times 60\,\mathrm{cm} \times 15\,\mathrm{cm}$, separate adjacent layers of PRCs. Above layer-3, 13 layers of concrete blocks are arranged in an inverted pyramid shape with a $45^\circ$ angle. These layers provide a total grammage of approximately $550\,\mathrm{g}\,\mathrm{cm}^{-2}$ above layer-0, with an energy threshold of $1\,\mathrm{GeV} $ for vertical muons\cite{hayashi2005} and act like an absorber for electromagnetic component. 
 \section{Motivation}
Separating muons from hadrons at G3MT is essential because the concrete grammage above each module cannot completely absorb the hadronic component of the air shower. Since hadrons are charged particles, they also produce signals in the detector that can be falsely interpreted as muonic signals. In earlier analyses, to solve this problem, only air showers with a core located atleast $60\,\mathrm{m}$ radially away from the centre of G3MT were considered~\cite{varsi2023}, as most hadrons are concentrated near the shower core. However, this approach is not efficient, leading to a significant loss of shower statistics for events near the detector. Therefore, a more precise separation method of muon and hadron components is needed, so that the muonic signals can be confidently identified, enabling a larger usable dataset. Furthermore, if the number of hadrons in a shower can be reliably estimated, this information could serve as an additional parameter for studying other properties of the air shower. Machine learning algorithms can be highly effective in distinguishing muonic secondaries from hadronic secondaries hitting G3MT. 

\section{Single particle classification}
Starting with a simpler case, we wanted to test the capabilities of various machine learning algorithms in identifying muons and hadrons striking one at a time on G3MT. For this purpose Geant4 simulations of G3MT were used. Within energy $100-158\mathrm{TeV}$ of the primary proton, we could generate 8479 muon and 2252 hadron events. Different machine learning algorithms like Decision Trees, RandomForest, XGBoost and Neural Networks were used to perform this task of identification. The input features were scalar, derived from array features like Station, Module, Layer, Counter, of length equalling the total number of hits in an event. The training was performed on $1000$ events of each muons and hadrons and rest of the events were used for testing purpose. The following results were obtained for the same.
\begin{table}[H]
\centering
\begin{tabular}{|c|c|c|c|c|}
\hline
\# & \textbf{NN} & \textbf{DT} & \textbf{RF} & \textbf{XGB} \\ \hline
\textbf{Muons identified as Muons} & 5854     & 6137     & 6588      & 6652     \\ 
\textbf{Muons identified as Hadrons}      & 1625      & 1342      & 891      & 827     \\ 
\textbf{Hadrons identified as Hadrons}     & 1126      & 1085      & 1062      & 1101     \\ 
\textbf{Hadrons identified as Muons}      & 126      & 167      & 190      & 151      \\ \hline
\textbf{Accuracy} & \textbf{80\%} & \textbf{82.7\%} & \textbf{87.6\%} & \textbf{88.7\%} \\ \hline
\textbf{Muon Signal loss} & \textbf{21.7\%} & \textbf{17.9\%} & \textbf{11.9\%} & \textbf{11.1\%} \\ \hline
\textbf{Hadron fake rate} & \textbf{10.06\%} & \textbf{13.3\%} & \textbf{15.2\%} & \textbf{12.1\%} \\ \hline
\end{tabular}
\caption{Confusion Matrix for the test set for the four machine learning algorithms. The following abbreviations are used- NN: Neural Network, DT: Decision Trees, RF: Random Forests, XGB: XGBoost.}
\label{confusion_table}
\end{table}
The ROC curve for the classification model is plotted below.
\begin{figure}[H]
    \centering
    \begin{overpic}[width=0.7\linewidth]{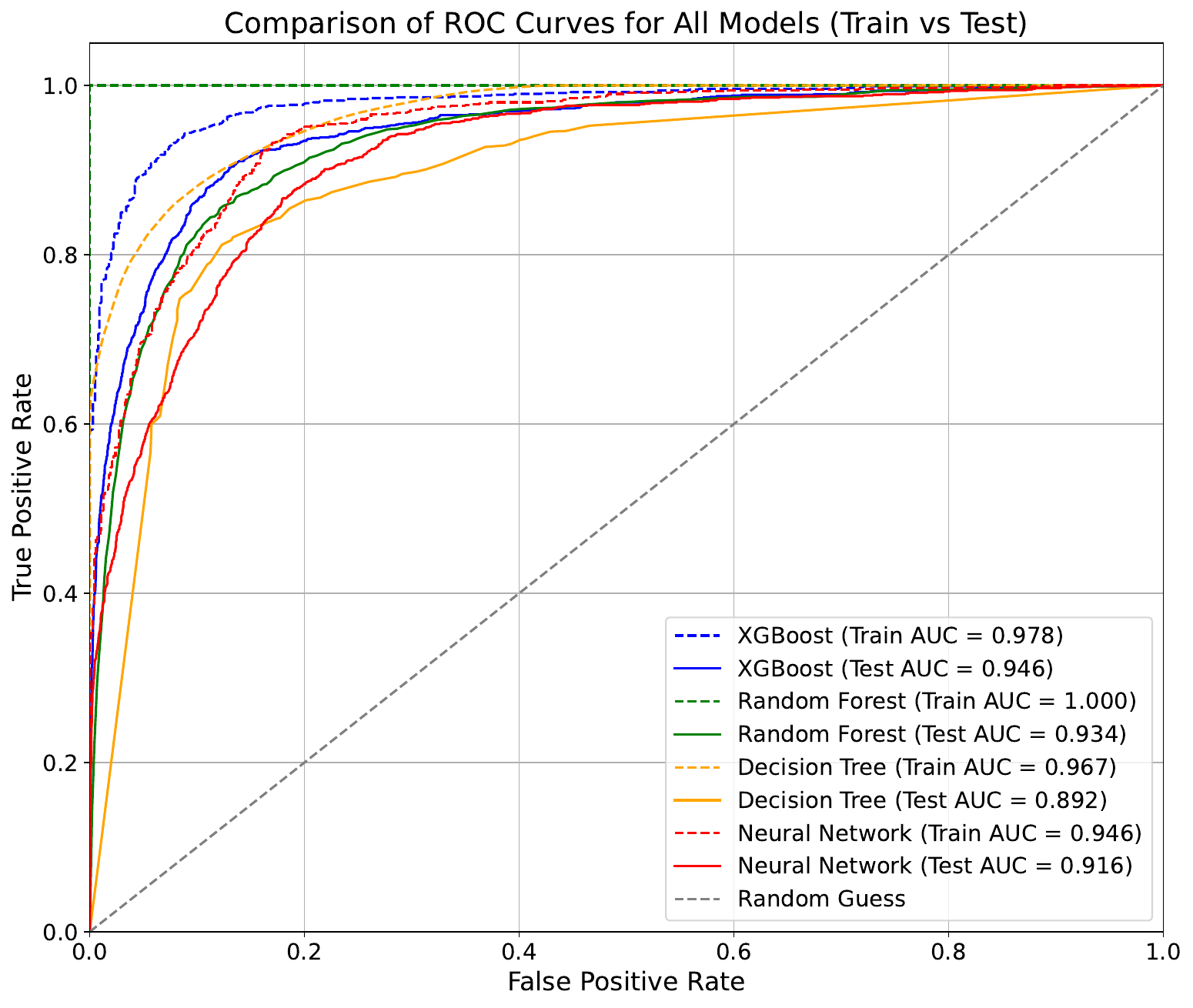}
        \put(70,70){\color{red} Preliminary}
    \end{overpic}
    \caption{ROC curve for each algorithm used in identifying single incoming muons and hadrons}
    \label{fig:enter-label}
\end{figure}
 
In classifying single particles, XGBoost emerges as the best algorithm, achieving the highest accuracy (88.7\%) and the lowest muon signal loss (11.1\%), making it the most reliable option overall.
\section{Graph node classification for multiple particles}
Now to identify many particles striking G3MT simultaneously, we started by randomly sampling single-particle events to create multiparticle events. For each module, $2000$ events with $1-10$ total particles (muons and hadrons both) were selected. Each event was modelled as a graph, with each hit of proportional counters as a node. Since each hit is a node, 11 input features were used namely, Station, Module, Layer, Counter, xmin, xmax, ymin, ymax, zmin, zmax and deposited energy dE. 
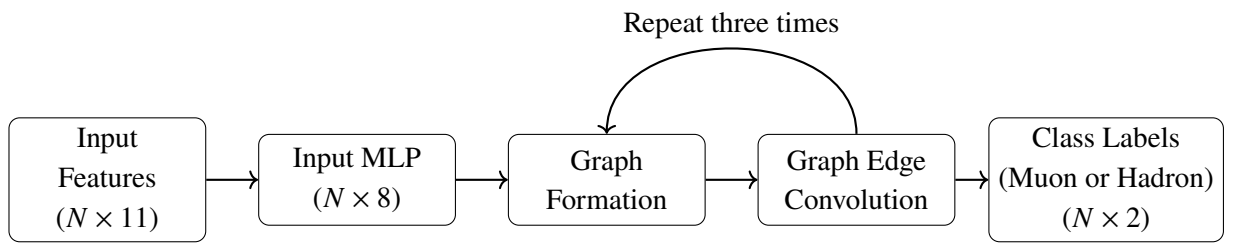
\begin{figure}[H]
    \centering
    \begin{tikzpicture}[node distance=3.5cm and 3.3cm, on grid, auto]

    \tikzstyle{layer} = [rectangle, draw=black, rounded corners, minimum height=1.2cm, minimum width=2.6cm, text centered, align=center]
     \tikzstyle{arrow} = [->, thick]

    \node (inputx) [layer] {Input \\ Features \\ ($N \times 11$)};
    \node (inputmlp) [layer, right=of inputx] {Input MLP \\ ($N \times 8$)};
    \node (edge1) [layer, right=of inputmlp] {Graph \\Formation };
    \node (edge2) [layer, right=of edge1] {Graph Edge\\ Convolution};
    \node (probs) [layer, right=of edge2] {Class Labels \\ (Muon or Hadron) \\ ($N \times 2$)};

    \draw [arrow] (inputx) -- (inputmlp);
    \draw [arrow] (inputmlp) -- (edge1);
    \draw [arrow] (edge1) -- (edge2);
    \draw [arrow] (edge2) -- (probs);
    \draw[arrow]
    (edge2.north) 
    .. controls +(0,1.5) and +(0,1.5) ..
    (edge1.north)
    node[midway, above] {Repeat three times};

    
    \end{tikzpicture}
    \caption{Block diagram of Graph Neural Network architecture used for node classfication task.}
    \label{fig:gnn-architecture-loop-correct}
\end{figure}
We first performed graph node classification using edge convolution operation\cite{wang2019} to identify which particle the proportional counter hit belonged to, a muon or a hadron. For this purpose, $584000$ multiparticle events were generated, $20\%$ of which were used for training. The following is the classification report and ROC curve,
\begin{figure}[H]
\centering
\begin{minipage}{0.38\linewidth}
    \centering
    \begin{tabular}{lcccc}
    \toprule
    \textbf{Class} & \textbf{Precision} & \textbf{Recall} & \textbf{F1-score} & \textbf{Support} \\
    \midrule
    0 & 0.83 & 0.86 & 0.85 & 15{,}421{,}545 \\
    1 & 0.92 & 0.89 & 0.90 & 24{,}111{,}413 \\
    \midrule
    \textbf{Accuracy} & \multicolumn{3}{c}{0.88} & 39{,}532{,}958 \\
    \textbf{Macro Avg} & 0.87 & 0.87 & 0.87 & 39{,}532{,}958 \\
    \textbf{Weighted Avg} & 0.88 & 0.88 & 0.88 & 39{,}532{,}958 \\
    \bottomrule
    \end{tabular}
    \captionof{table}{Classification report for graph node classification on test set.}
    \label{tab:node-class-report}
\end{minipage}
\hfill
\begin{minipage}{0.50\linewidth}
    \centering
    \begin{overpic}[width=0.95\linewidth]{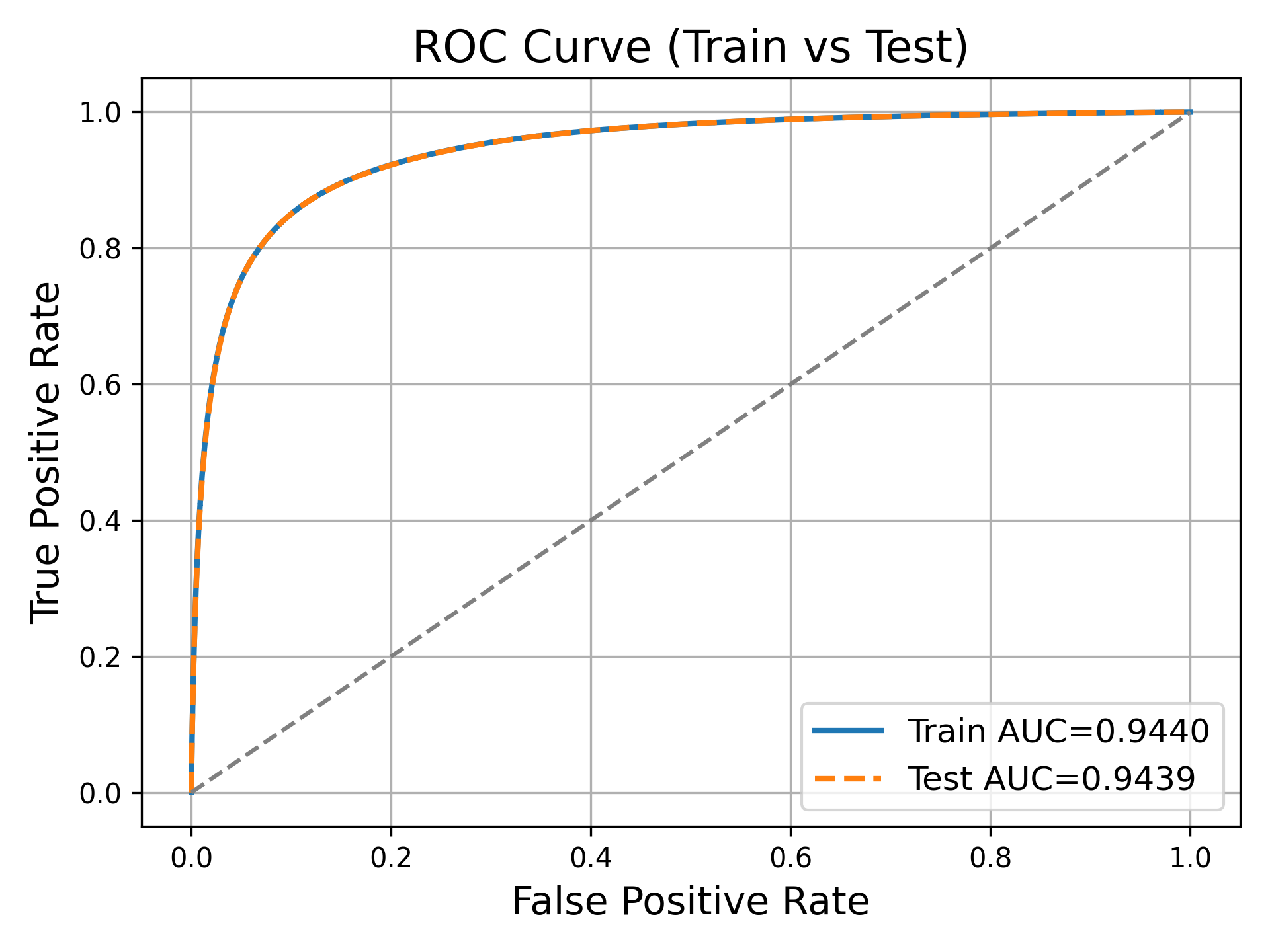}
        \put(60,60){\color{red} Preliminary}
    \end{overpic}
    \caption{ROC curve for Node Classification}
    \label{fig:roc-node-class}
\end{minipage}
\end{figure}

\section{Dynamic Reduction Network for Regression}
Now, to identify the number of particles and muons in an event, a regression model was trained using a dynamic reduction network \cite{aamir2024}. It is built on top of the graph node classification model. The output of the previous model along with the same 11 input features were used as input to DRN. The following diagram depicts the DRN model's architecture.

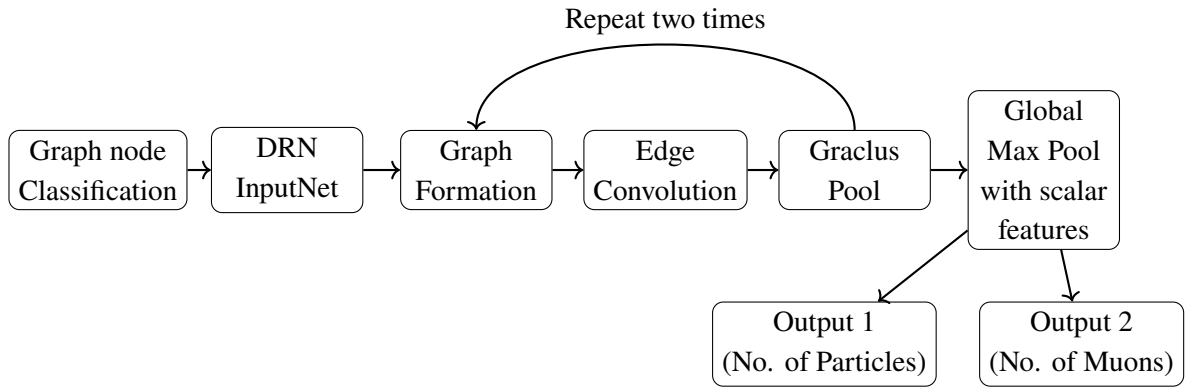
\begin{figure}[H]
\centering
\begin{tikzpicture}[
    node distance=1cm and 2.5cm,
    on grid,
    auto
]

\tikzstyle{block} = [
    draw,
    rectangle,
    rounded corners,
    minimum height=1cm,
    minimum width=2cm,
    align=center
]
\tikzstyle{arrow} = [->, thick]

\node (particlegnn) [block] {Graph node \\ Classification};
\node (drn_inputnet) [block, right=of particlegnn] {DRN \\ InputNet};
\node (graph) [block, right=of drn_inputnet] {Graph \\ Formation};
\node (edge) [block, right=of graph] {Edge \\Convolution};
\node (graclus) [block, right=of edge] {Graclus \\ Pool};
\node (globalpool) [block, right=of graclus] {Global \\ Max Pool\\
with scalar \\features};

\node (output1) [block, below left=2.3cm and 2.9cm of globalpool] {Output 1 \\ (No. of Particles)};
\node (output2) [block, below right=2.3cm and 0.5cm of globalpool] {Output 2 \\ (No. of Muons)};

\draw[arrow] (particlegnn) -- (drn_inputnet);
\draw[arrow] (drn_inputnet) -- (graph);
\draw[arrow] (graph) -- (edge);
\draw[arrow] (edge) -- (graclus);
\draw[arrow] (graclus) -- (globalpool);
\draw[arrow] (globalpool) -- (output1);
\draw[arrow] (globalpool) -- (output2);

\draw[arrow]    
    (graclus.north) 
    .. controls +(0,1.5) and +(0,1.5) ..
    (graph.north)
    node[midway, above, sloped] {Repeat two times};

\end{tikzpicture}
\caption{Block diagram of the dynamic reduction network architecture for regression task.}
\label{fig:drn-architecture}
\end{figure}
Some scalar features were derived from the 11 array features and combined with pooled layers of edge convolution for two regression output MLP for total number of particles and muons. Again the same 584000 events were used for this regression task, $20\%$ of which were used for training.

The following table tells various metrics obtained for the
DRN model on the test set, where MAE refers to mean absolute error, MSE to mean square error and $R^2$ to the R squared score.
\begin{table}[H]
\centering
\begin{tabular}{lccc}
\hline
\textbf{Metric} & \textbf{Number of particles} & \textbf{Number of muons} \\
\hline
MAE & 0.809 & 0.746 \\
MSE & 1.118 & 0.963 \\
$R^2$ & $\mathbf{0.837}$ & $\mathbf{0.866}$ \\
\hline
\end{tabular}
\caption{Test metrics for no\_of\_particles and no\_of\_muons predictions.}
\label{tab:drn_test_metrics}
\end{table}
The following graphs depict the relationship between predicted and true number of particles and muons respectively. 
\begin{figure}[H]
    \centering
    \begin{minipage}{0.48\textwidth}
        \centering
        \begin{overpic}[width=\linewidth]{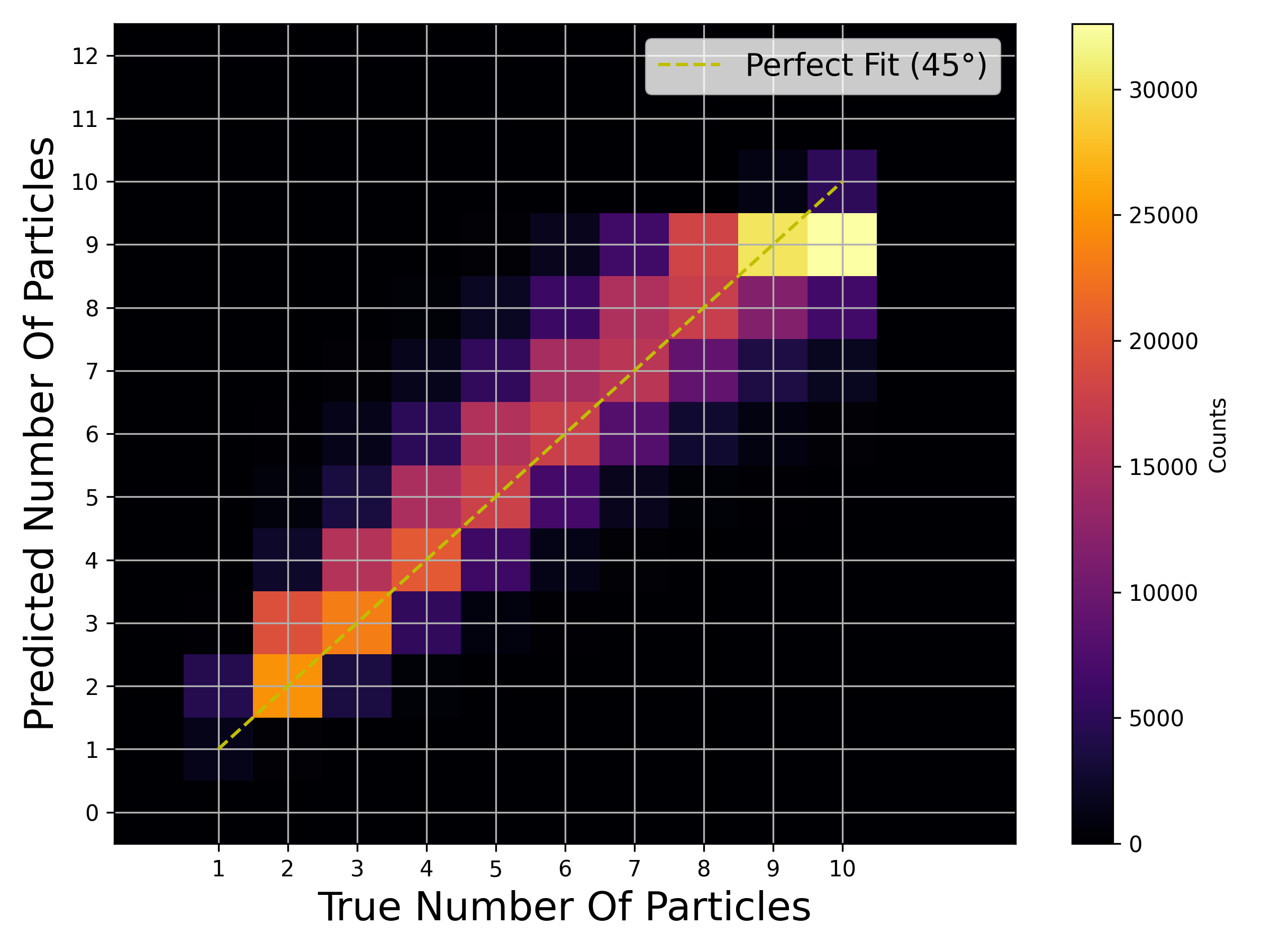}
        \put(20,60){\color{red} Preliminary}
        \end{overpic}
        \caption{2D histogram of Predicted vs True number of particles.}
        \label{fig:hist2d_particles}
    \end{minipage}
    \hfill
    \begin{minipage}{0.48\textwidth}
        \centering
        \begin{overpic}[width=\linewidth]{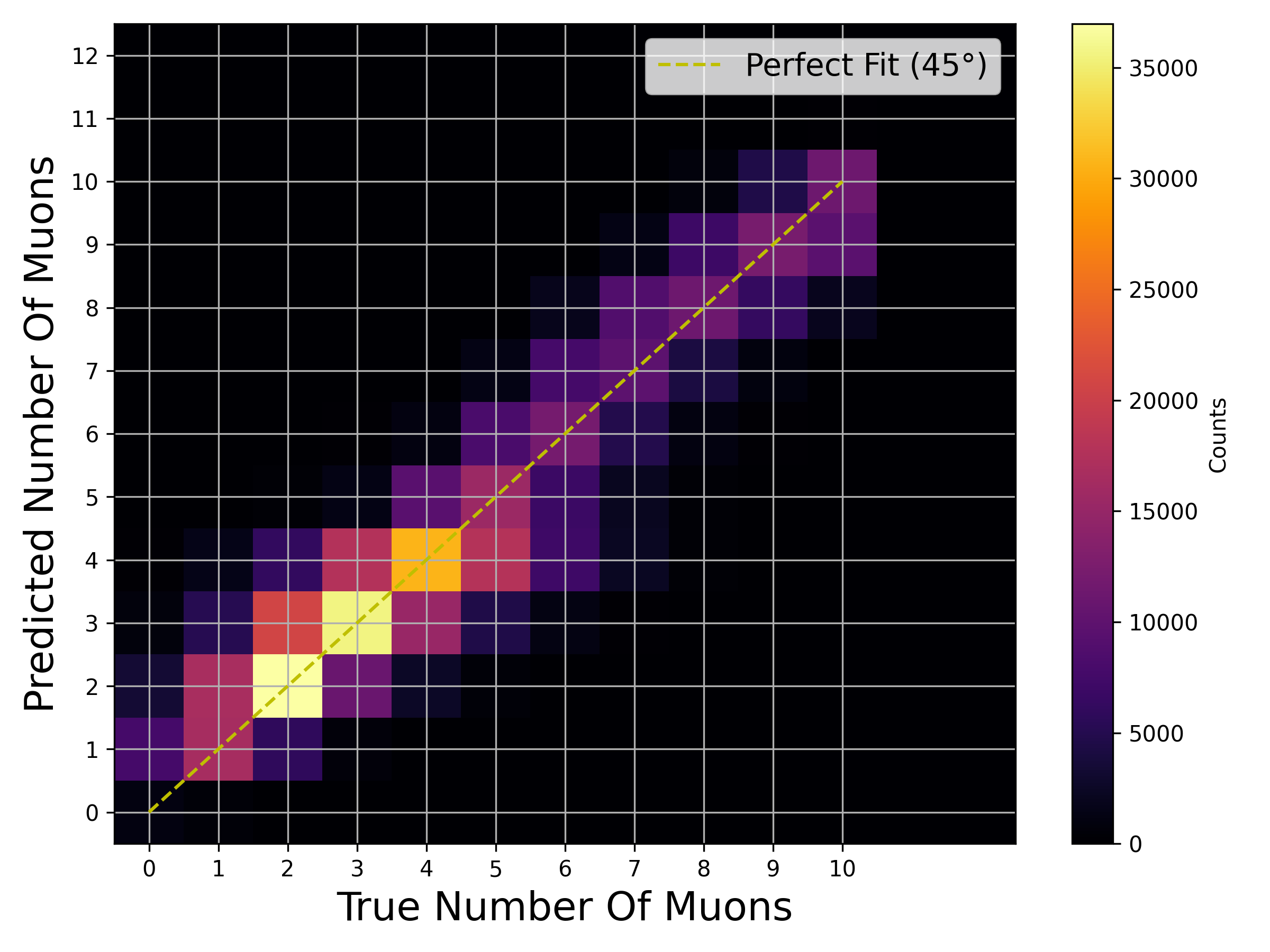}
        \put(20,60){\color{red} Preliminary}
        \end{overpic}
        \caption{2D histogram of Predicted vs True number of muons.}
        \label{fig:hist2d_muons}
    \end{minipage}
\end{figure}
The below graphs shows the resolution plot for both the predicted numbers.
\begin{figure}[H]
    \centering
    \begin{minipage}{0.48\textwidth}
        \centering
        \begin{overpic}[width=\linewidth]{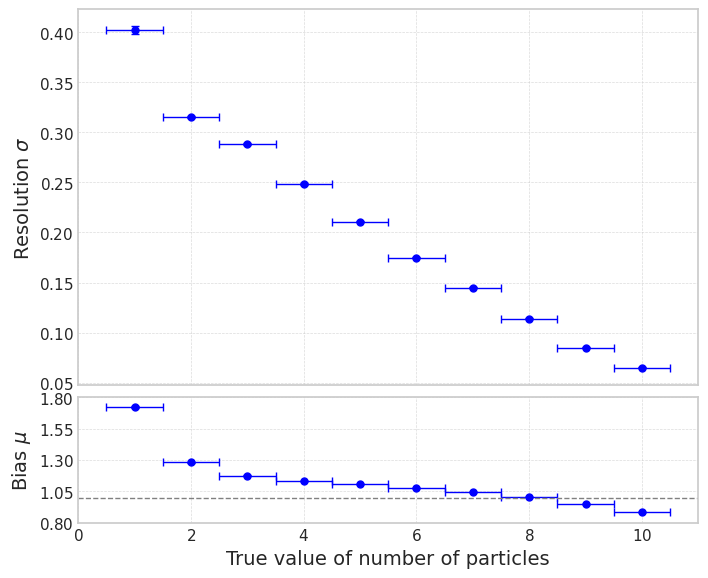}
        \put(60,60){\color{red} Preliminary}
        \end{overpic}
        \caption{Resolution and bias plot for the total number of particles.}
        \label{fig:hist2d_particles}
    \end{minipage}
    \hfill
    \begin{minipage}{0.48\textwidth}
        \centering
        \begin{overpic}[width=\linewidth]{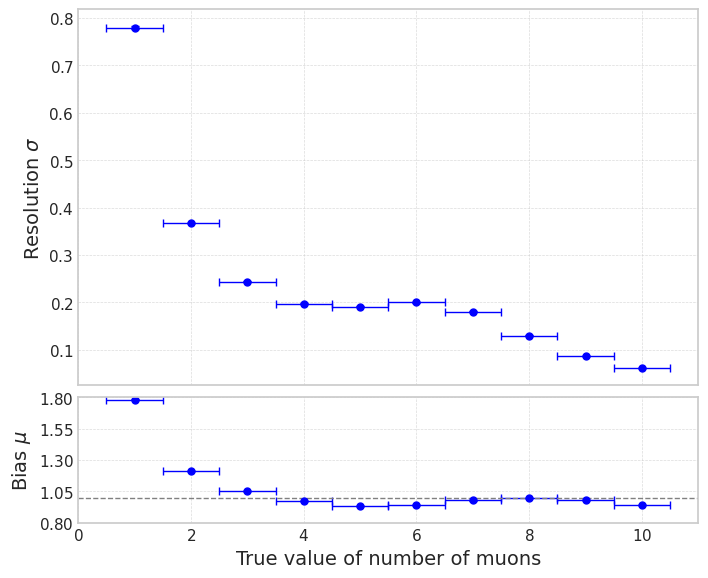}
        \put(60,60){\color{red} Preliminary}
        \end{overpic}
        \caption{Resolution and bias plot for the total number of muons.}
        \label{fig:hist2d_muons}
    \end{minipage}
\end{figure}
In our work, \textbf{bias} is defined as the mean ratio between the predicted and true values in each bin and \textbf{Resolution} is the standard deviation of the relative residuals, describing the spread of predictions around the true value as a fraction of that value.
\[
\text{Bias} = \left\langle \frac{\hat{x}}{x_\text{true}} \right\rangle  \text{ ; }
\text{Resolution}
   = \operatorname{Std}\!\left( \frac{\hat{x}-x_\text{true}}{x_\text{true}} \right) .
\]

\section{Conclusion}
This work tells us that machine learning algorithms are effective in distinguishing muons and hadrons. We saw that for single particle classification, XGBoost emerged to be the best algorithm with an accuracy score of $88.7\%$. Then for multiparticle events we performed graph node classification using edge convolution operation and achieved an accuracy score of $88\%$. A dynamic reduction network model was built on top of the node classification model to regress the total number of particles and muons. The DRN model achieved an $R^2$ score of $0.837$ and $0.866$ with mean absolute error of $0.809$ and $0.746$ for total number of particles and muons, respectively.

Overall, these results show that graph-based machine learning methods can reliably estimate the particle content of air showers, even when many particles arrive together and their hits overlap in the muon detector.

\end{document}